\documentclass[12pt,a4paper,reqno]{amsart}
\usepackage{graphicx}
\usepackage{amsmath,amssymb,amsthm,array}
\usepackage{amsfonts, amsmath, amsthm, amssymb}
\usepackage{hyperref}
\newcommand{\sech}{\mathop{\mathrm{sech}}\nolimits}%

\begin{document}
\title[Soliton transmutations in KdV---Burgers layered media]{SOLITON TRANSMUTATIONS IN KDV---BURGERS LAYERED MEDIA}

\author{Alexey Samokhin}\vspace{6pt}

\address{Institute of Control Science, Russian Academy of Sciences}\vspace{6pt}

\email{ samohinalexey@gmail.com}\vspace{6pt}

\begin{abstract}
We study the behavior of the soliton  which, while moving in non-dissipative medium encounters a barrier with dissipation. The modelling included the case of a finite dissipative layer as well as  a wave passing from a dissipative layer into a non-dissipative one and vice versa.
New effects are presented in the case of numerically finite barrier on the soliton path: first, if the form of dissipation distribution has a form of a frozen soliton, the wave that leaves the dissipative barrier becomes a bi-soliton and a reflection wave arises as a comparatively  small and quasi-harmonic oscillation.  Second, if the dissipation is negative (the wave, instead of loosing energy, is pumped with it) the passed wave is a soliton of a greater amplitude and velocity. Third, when the travelling wave solution of the KdV-Burgers (it is a shock wave in a dissipative region) enters a non-dissipative layer this shock transforms into a quasi-harmonic oscillation known for the KdV. \vspace{1mm}

\end{abstract}

\maketitle

\section{Introduction}

 The behavior of solutions to the KdV - Burgers equation is a subject of various recent research, \cite{key-1}--\cite{key-6}; the present paper is a continuation of \cite{key-2}.

 The generalized KdV-Burgers (KdV-B) equation
 \begin{equation}%\label{01}
    u_t(x,t)=\gamma(x)u_{xx}(x,t)+ 2u(x,t)u_x(x,t)+ u_{xxx}(x,t).
    \end{equation}%\vspace{1mm}

 It is related to the viscous and dispersive medium. The layered media consist of layers with both dispersion an dissipation (modelled by KdV-B) and layers  without dissipation (KdV).

\begin{itemize}
  \item $\gamma(x)=0\Rightarrow$   KdV $\Rightarrow$ travelling waves solutions (TWS) are solitons;
  \item $\gamma(x)=\mathrm{const}>0\Rightarrow$  KdV-Burgers $\Rightarrow$ TWS are shock waves.
\end{itemize}\vspace{3mm}

Thus we consider the following possibilities for $\gamma(x)$.

\begin{enumerate}
  \item The two-layer case: $\gamma(x)=\alpha(1-\theta(x))$, $\theta(x)=\mathrm{sign}(x)$ --- the Heaviside step function; or $\gamma(x)=\alpha(1-\tanh(x))$, its smooth analog;
  \item The three-layer case: $\gamma(x)=\alpha(\theta(x-\beta)-\theta(x+\beta))$, a $\Pi$-form density of viscosity, or $\gamma(x)=\alpha\sech^{2}(\beta x)$, its smooth analog with (numerically) compact support.
\end{enumerate}

 Initial value-boundary problem  for the KdV-Burgers :
 \begin{equation}%\label{08}
u(x,0) =u_0(x,a,s)=6a^2 \sech^{2}(a(x+s)), \;  u_x(\pm \infty,t) =0:
\end{equation}

 This is the initial placement of a soliton

 $u_0(x,a,s)=\left.6a^2 \sech^{2}(4a^3t+a(x+s))\right|_{t=0}$ moving to the left.

 For numerical computations  $ x\in[a,b]$ for appropriately large $a,\;b$.\vspace*{10mm}

\section{Three layers case}

\subsection{$\Pi$ - type dissipation barrier}

This case models a passage from non-dissipative half-space to another  one passing through a dissipative layer (a process similar to a wave passing through an air-glass-air pile). We take $\gamma(x)=\alpha(\theta(x-\beta)-\theta(x+\beta))$ as a $\Pi$-form density of viscosity (the three-layer case) distribution function to present the layer separating these half-spaces.

Our experiments show that the initial soliton  behaves as the one of the KdV at the right half-space and as a diminished soliton or a bi-soliton at the left one.

\begin{figure}[h]
  \begin{minipage}{11pc}
\includegraphics[width=11pc]{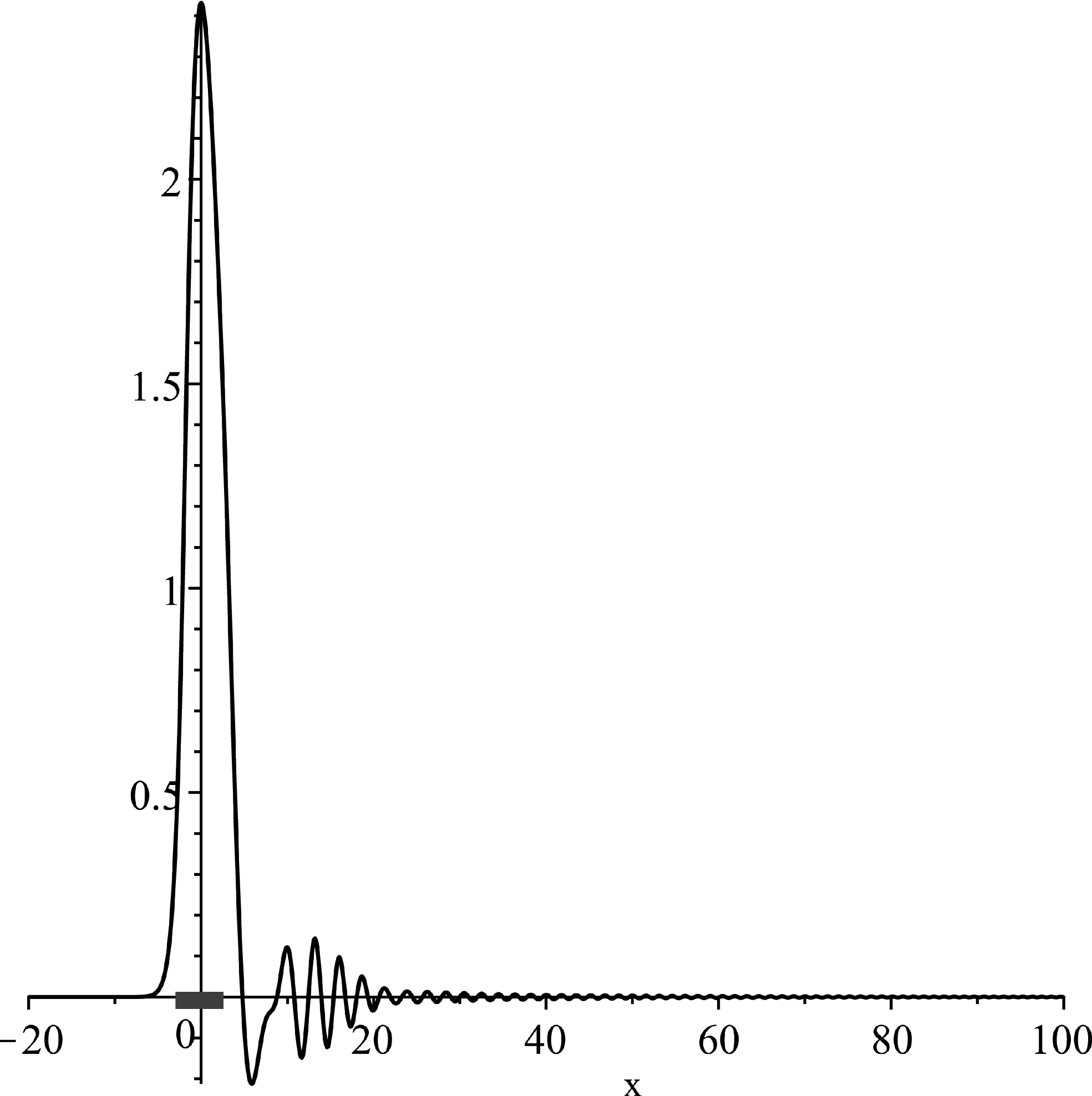}
\end{minipage}
 \begin{minipage}{11pc}
\includegraphics[width=11pc]{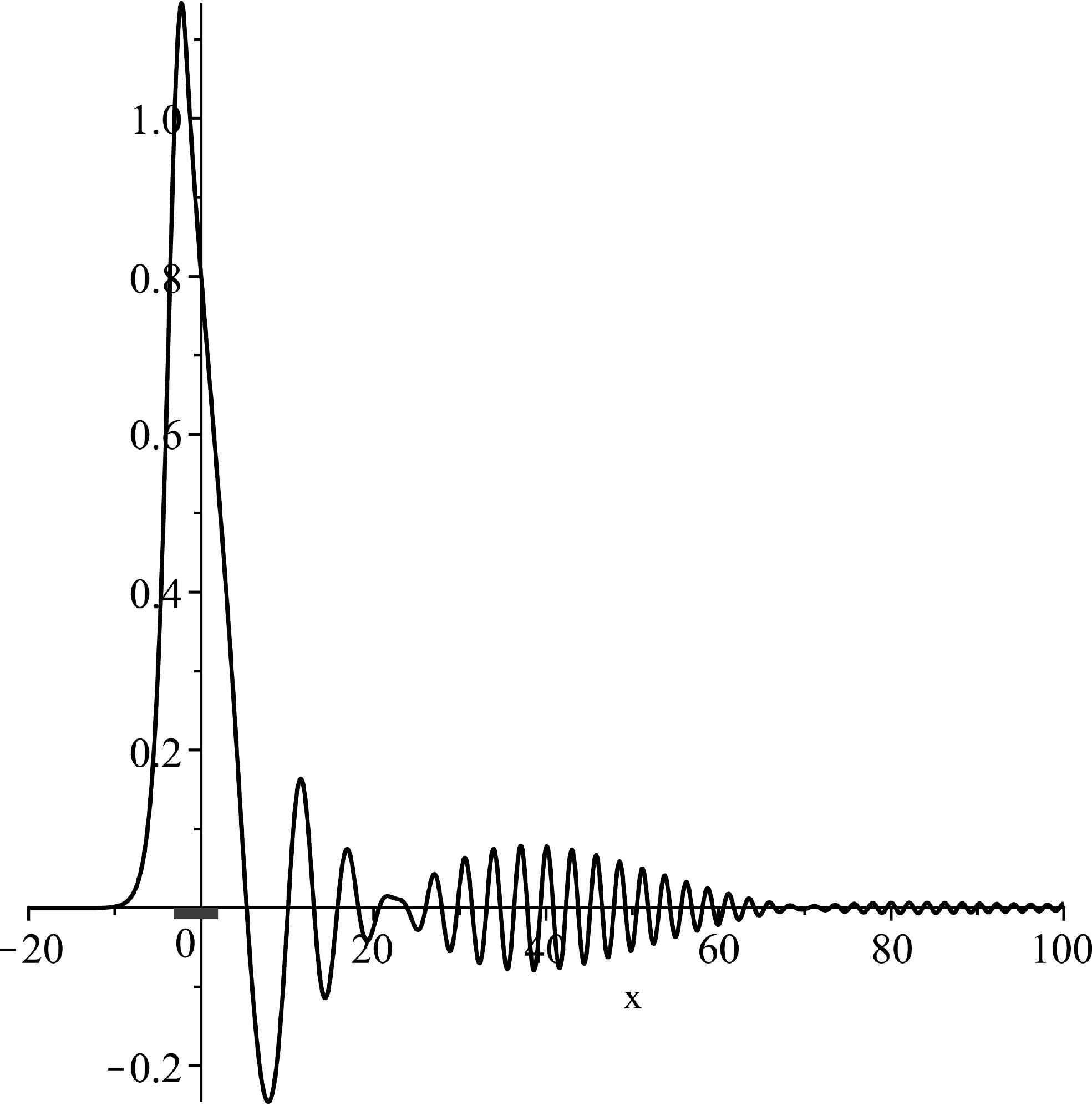}
\end{minipage}
\caption{  \textbf{Left: }Soliton ($a=1$)  passing a thin viscose $\Pi$-type  layer ($\alpha=2.5$, $\beta=2$); $t=1$
\protect\newline \hspace*{1.25cm}\textbf{Right: } $t=3$}\label{ST3}
\end{figure}

The process of transition  is natural enough. The transient wave in a dissipative media looses energy and speed to become a lesser and slower solution at the left non-dissipative half-space; and a reflected wave is seen at the right half-space as it is shown at  figures \ref{ST3}, \ref{ST4}. You may also see the \textit{\textbf{13.avi }} Maple-generated movie attached to this paper.

\begin{figure}[h]
\begin{minipage}{11pc}
\includegraphics[width=11pc]{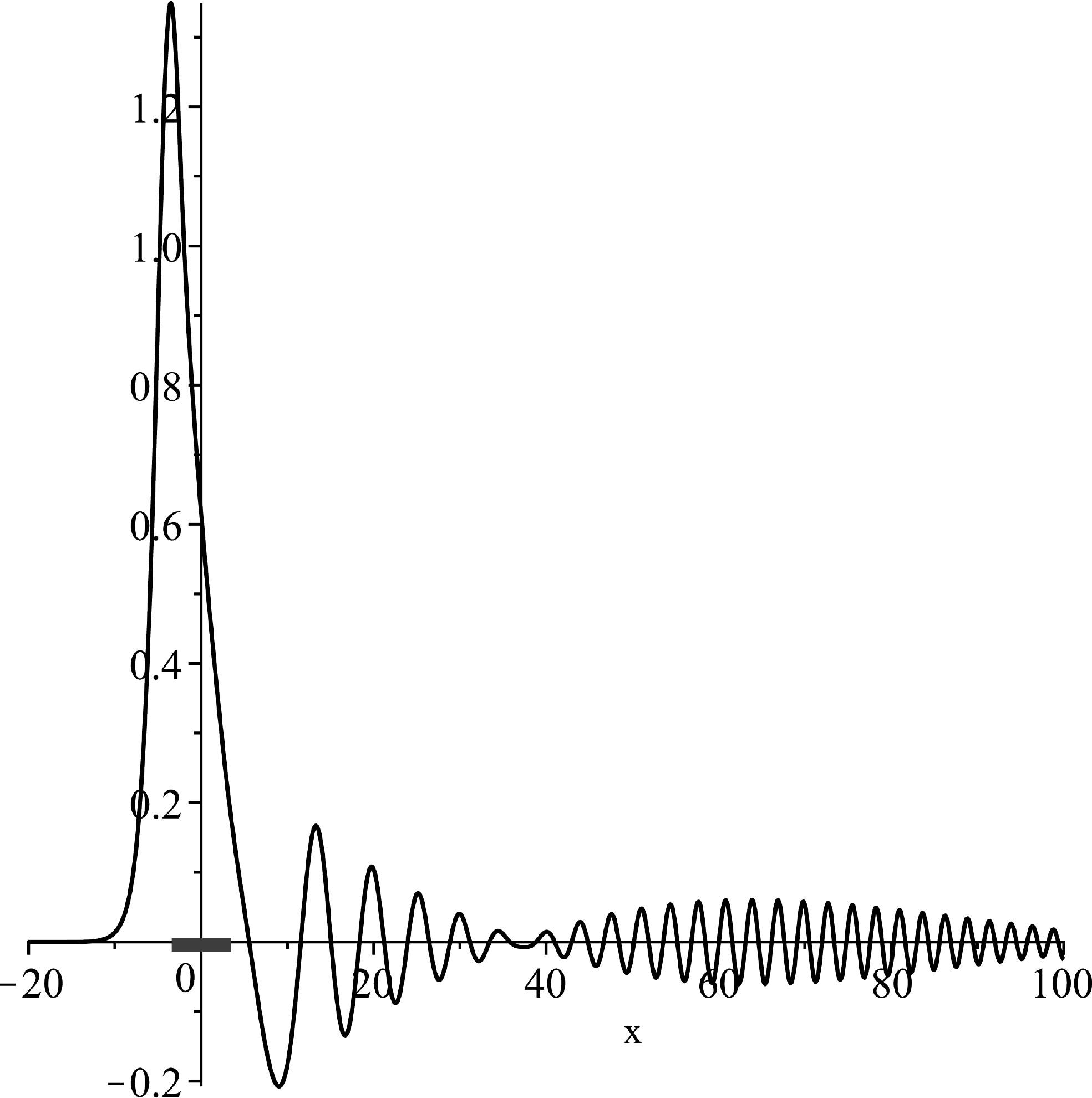}
\end{minipage}\hspace{2pc}%
\begin{minipage}{11pc}
\includegraphics[width=11pc]{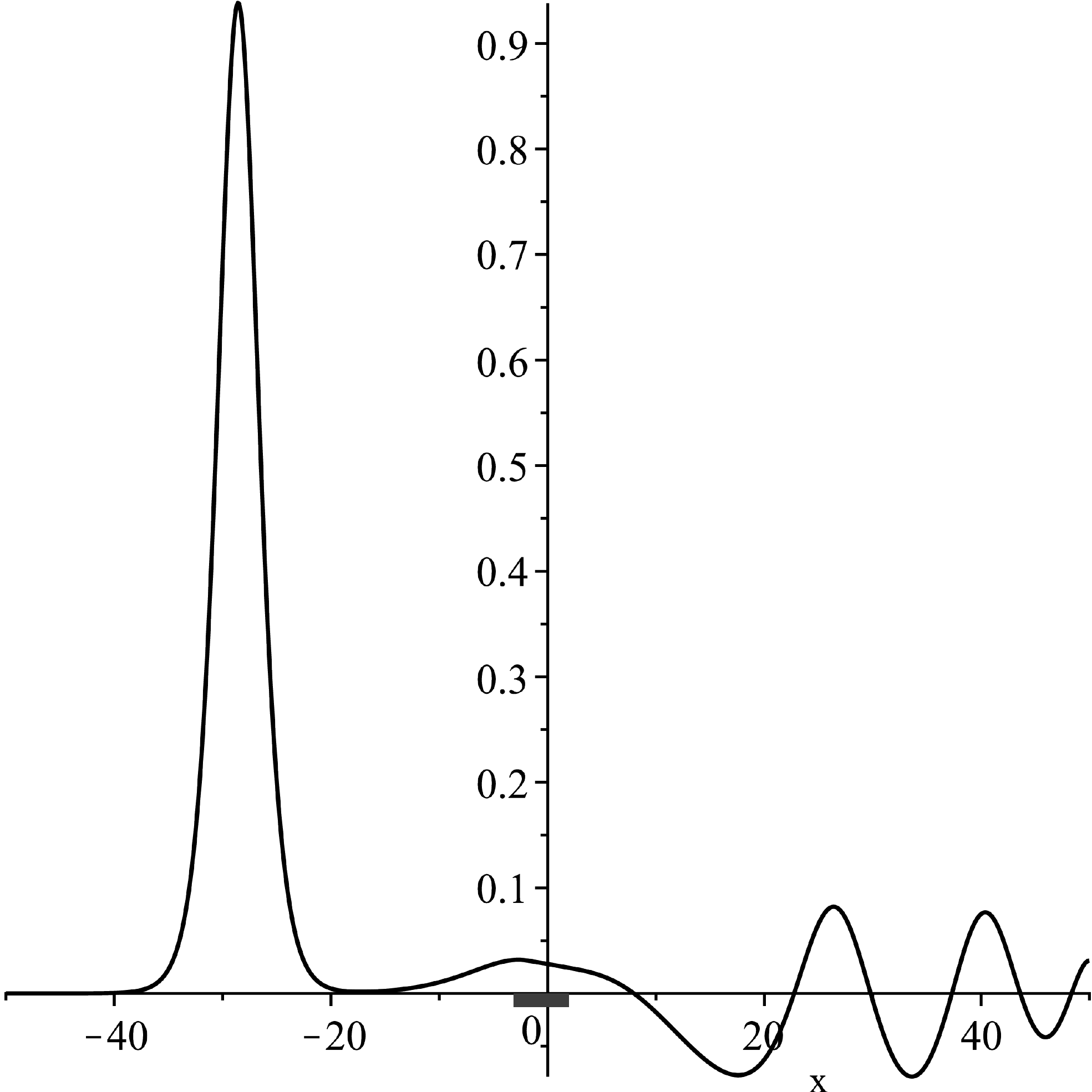}
\end{minipage}
\caption{\label{ST4}  \textbf{Left: }Soliton ($a=1$)  passing a thin viscose $\Pi$-type  layer ($\alpha=2.5$, $\beta=2$); $t=5$. \protect\newline
\hspace*{1.25cm}\textbf{Right: }Soliton (initially $a=1$) becomes that of $a=0.4$ after passing a thin viscose $\Pi$-type  layer ($\alpha=6$, $\beta=2$); $t=50$.}
\end{figure}

The size of the reflected breather is connected, in particular, to the properties of the barrier $\alpha$ and $\beta$. So it may be of a practical use: for instance, measuring it one can judge whether the layer connecting two details is uniform at different points.

\begin{figure}[h]
\begin{minipage}{13pc}
\includegraphics[width=11pc]{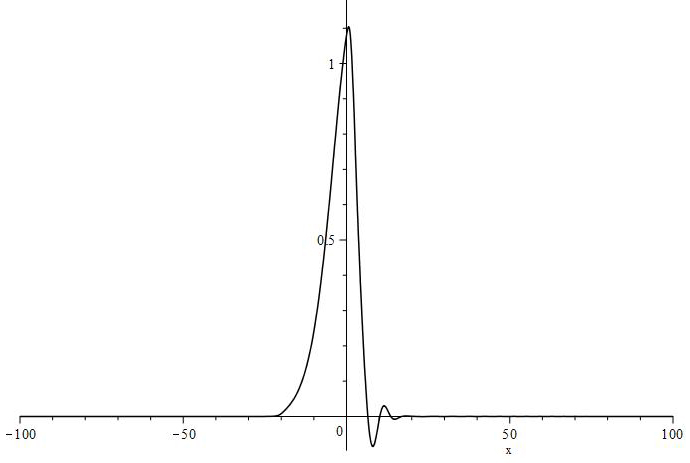}
\end{minipage}\hspace{2pc}%
\begin{minipage}{11pc}
\includegraphics[width=15pc]{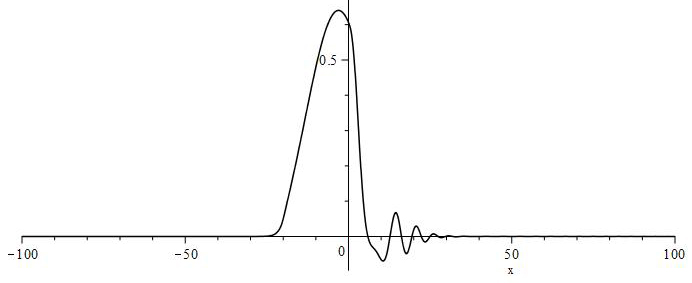}
\end{minipage}
\caption{\label{ST4}  \textbf{Left: }Soliton ($a=0.5$) passing $\Pi$-type obstacle located at: \textbf{$\mathbf{-20<x<0}$}, $t=7$ \protect\newline
\hspace*{1.25cm}\textbf{Right: } $t=10$.}\label{12}
\end{figure}

\begin{figure}[h]
\begin{minipage}{11pc}
\includegraphics[width=13pc]{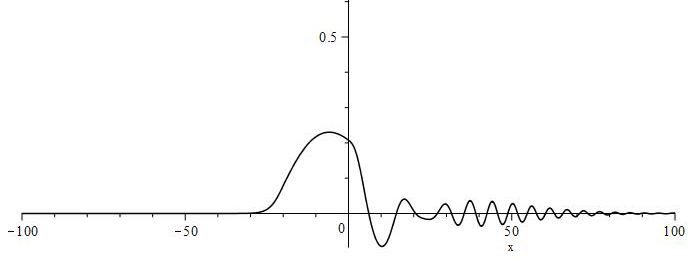}
\end{minipage}\hspace{2pc}%
\begin{minipage}{11pc}
\includegraphics[width=13pc]{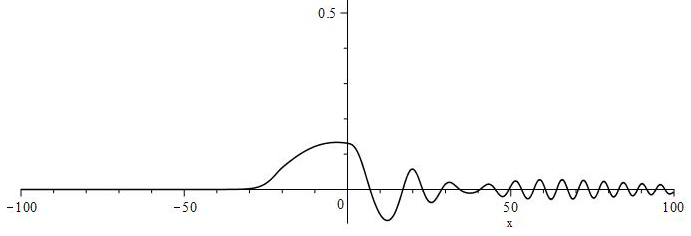}
\end{minipage}
\caption{\label{ST4}  \textbf{Left: }Soliton ($a=0.5$) passing $\Pi$-type obstacle located at: \textbf{$\mathbf{-20<x<0}$}, $t=20$ \protect\newline
\hspace*{1.25cm}\textbf{Right: } $t=30$.}\label{14}
\end{figure}

The natural question: is it possible to shut a soliton altogether using the dissipative layer? The absolute filter is impossible, but... see the  figures \ref{12}, \ref{14}. You may also see the \textit{\textbf{widePi.avi }} Maple-generated  movie attached to this paper.

\subsection{Soliton-form barrier}

\begin{figure}[h]
\includegraphics[width=18pc]{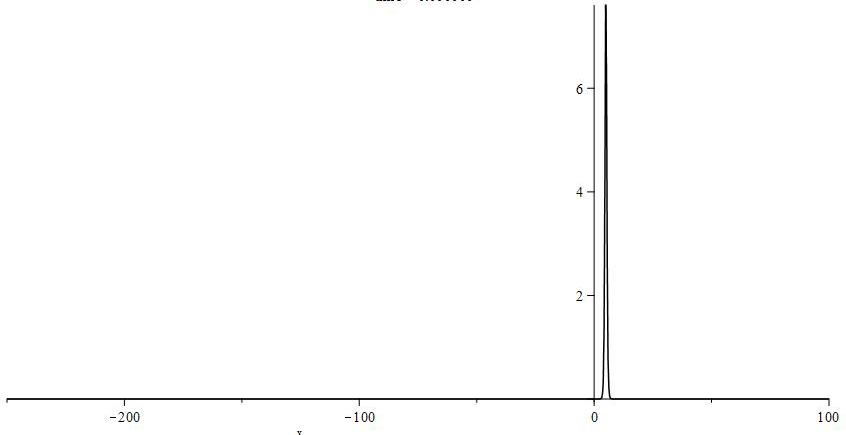}
\caption{ Soliton ($a=4$) passing a  thin viscose layer $\gamma(x)=2\sech^{2}(x)$, $t=0$ }\label{SL1}
\end{figure}

\begin{figure}[h]
\includegraphics[width=26pc]{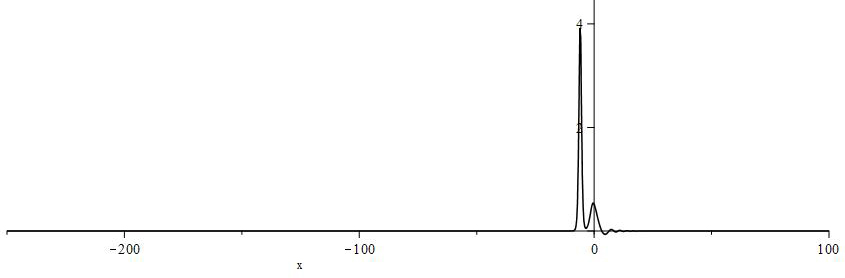}
\caption{\textsl{ Soliton ($a=4$)  passing a thin viscose layer $2\sech^{2}(x)$, $t=1$.}} \label{3}
\end{figure}

\begin{figure}[h]

\includegraphics[width=26pc]{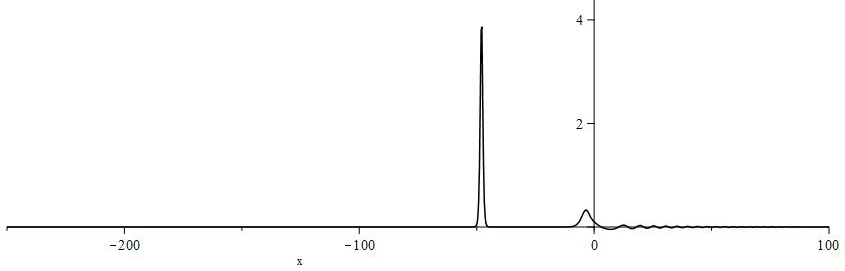}
\caption{ Soliton ($a=4$) passing a thin viscose layer $\gamma(x)=2\sech^{2}(x)$, $t=5$}\label{2}
\end{figure}

\begin{figure}[h]
\includegraphics[width=26pc]{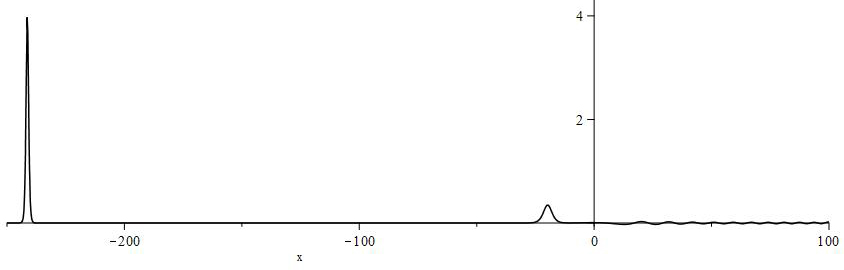}
\caption{\textsl{ Soliton ($a=4$)  passing a thin viscose layer $2\sech^{2}(x)$, $t=30$. The refraction coefficient for the first peak is $k\approx\frac{8}{4}\approx 2$}}\label{4}
\end{figure}

Since the barrier $\gamma(x)=\alpha\cosh^{-2}(\beta x)$ has a numerically compact support,  the transition process is similar to that for the $\Pi$-type obstacle as it is illustrated by figures \ref{SL1}--\ref{SL2}. You may also see the \textit{\textbf{bi-2.avi }} Maple-generated  movie attached to this paper.

For  pictures in this subsection the equation is chosen in a different but equivalent form

 \begin{equation}%\label{01}
    u_t(x,t)=\gamma(x)u_{xx}(x,t)+ 6u(x,t)u_x(x,t)+ u_{xxx}(x,t)
    \end{equation}%\vspace{1mm}

 with $\gamma(x)=2\sech^2(x)$\vspace{3mm}

 The solitons for the corresponding KdV,
$  u_t=+ 6u(x,t)u_x+ u_{xxx}$, are of the form $\frac 12 a^2\sech^2(\frac{a}{2}(a^2t+x)+s)$.
\vspace{3mm}

Note that $\gamma(x)$ has a form of the soliton for $a=2,\;s=t=0$ but  it is stationary, with zero velocity (a frozen soliton).
\vspace{3mm}

The transient wave becomes a bi-soliton --- this effect is not observed as distinctly for different type barriers. Both peaks have the right ratioo of heights to velocities.

\newpage

\newpage

\subsection{Pumping area instead of dissipationone:}

\begin{figure}[h]
\includegraphics[width=18pc]{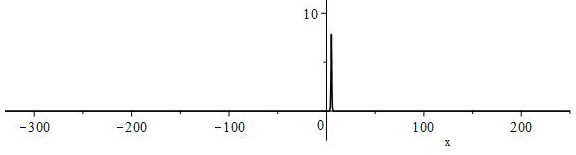}
\caption{ Soliton ($a=4$) at a thin negative-viscose layer $\gamma(x)=-2\sech^{2}(x)$, $t=0$} \label{SL2}
\end{figure}

\begin{figure}[h]
\includegraphics[width=24pc]{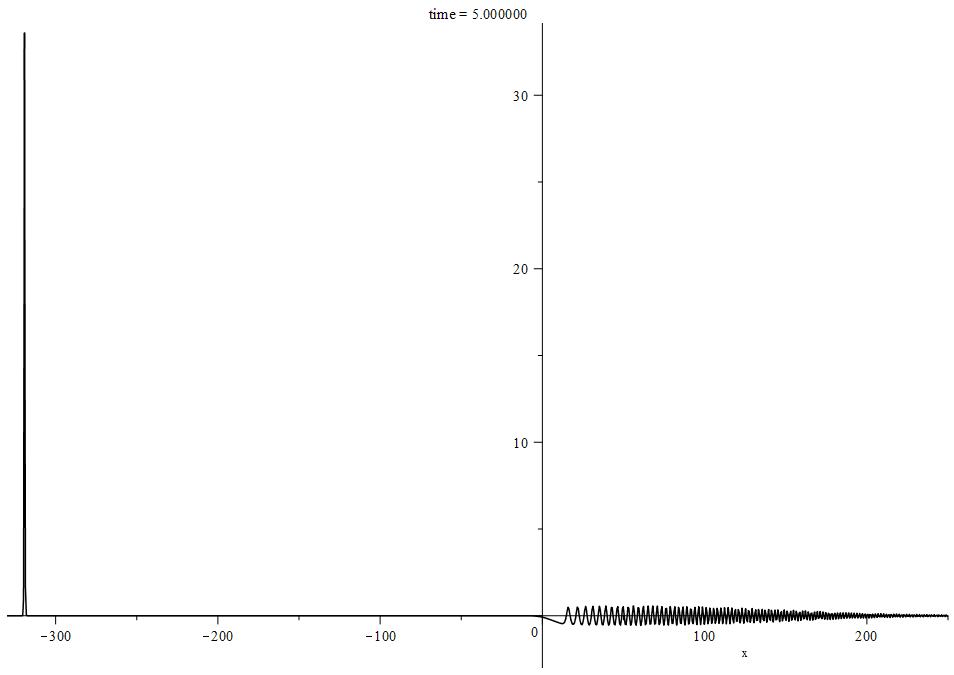}
\caption{ Soliton ($a=4$)  passing a thin negative-viscose layer $-2\sech^{2}(x)$, $t=5$.}\label{SL3}
\end{figure}

 Invert the sign of $\gamma$ and look at soliton ($a=4$)  crossing the "pumping" area ($\gamma=-2\sech^2(x)\ll 0$: the energy is not lost, but is acquired instead). The soliton comes out greater in amplitude and velocity; and the reflected wave gets a substantial impetus (figures \ref{SL2}--\ref{SL3}). You may also see the \textit{\textbf{bi-6.avi }} Maple-generated  movie attached to this paper.

\begin{figure}[h]
\includegraphics[width=24pc]{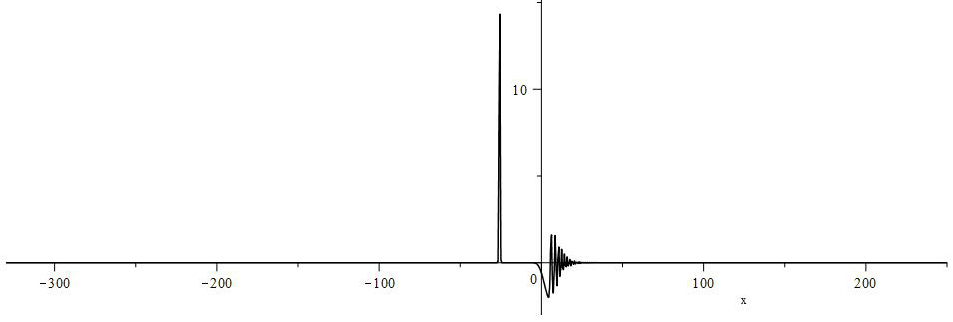}
\caption{ Soliton ($a=4$)  passing a thin negative-viscose layer $-2\sech^{2}(x)$, $t=0.7$. }
\end{figure}

\begin{figure}[h]
\includegraphics[width=24pc]{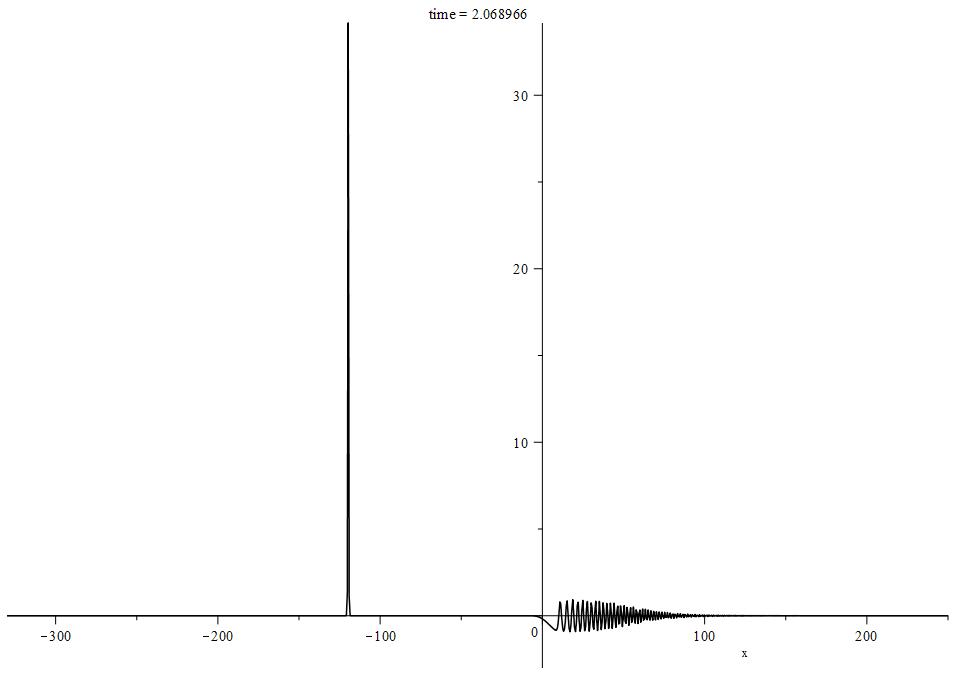}
\caption{ Soliton ($a=4$)  passing a thin negative-viscose layer $-2\sech^{2}(x)$, $t=2$.}
\end{figure}

\newpage

\section{Soliton in 2-layer medium}

This case models a passage from a dissipative half-space to a non-dissipative one. We take $\gamma(x)=0.5(1+\theta(x))$ as a  dissipation distribution function to present a single boundary separating these half-spaces.

Any localized solution behaves as the one of the KdV at the left half-space   and as a solution of KdV-B at the right one.

We modeled KdV-B shock wave entering a non-dissipative region:
the shock waves are TWS solutions for the KdV-B with $\gamma(x)=g=\mathrm{const},\; x>0 $. They have a form

\[u=
\frac{3}{50}g^2\sech^2(\frac{g}{10}(-Vt+s+x))+(\frac{3}{25}g\tanh(\frac{g}{10}(-Vt+s+x))-\frac{V}{2}
\]

If $u=0$ at $t\rightarrow -\infty$ is required, the sole such shock wave  is

\[
\frac{3}{50}g^2\sech^2(\frac{3}{125}g^3t+\frac{1}{10}gx+s)+\frac{3}{25}g^2\tanh((\frac{3}{125}g^3t+\frac{1}{10}gx+s)+\frac{3}{25}g^2
\]
\vspace{1mm}

As such a TWS moves from the right it passes the the stair boundary and enters the area without dissipation.

\begin{figure}[h]
\begin{minipage}{13.2pc}
\includegraphics[width=13.2pc]{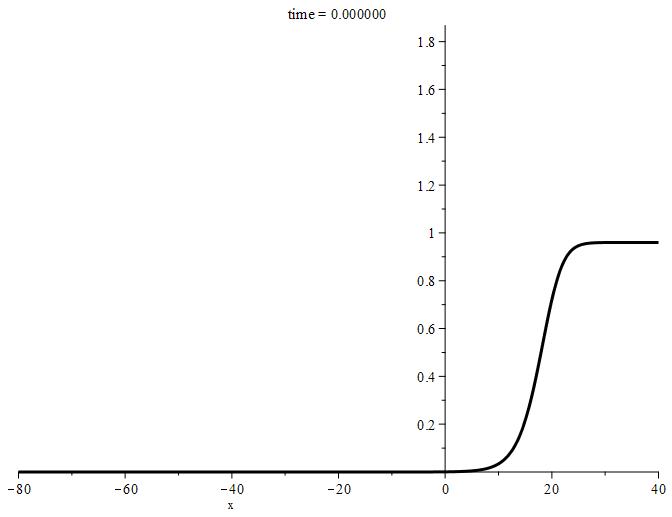}
\end{minipage}.
\begin{minipage}{13.2pc}
\includegraphics[width=13.2pc]{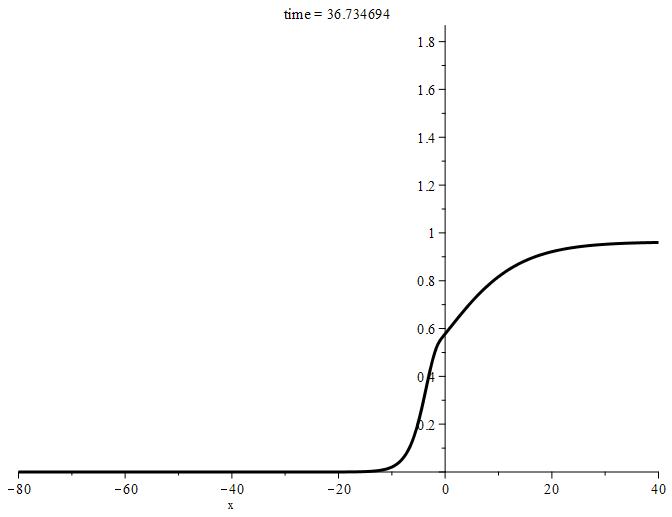}
\end{minipage}\caption{\textsl{\textbf{Left}:} Initial position of the shock wave $(6/25)\sech(-4+(1/5)x)^2+(12/25)\tanh(-4+(1/5)x)+12/25$, $t=0$. \protect\newline\textsl{\textbf{Right}:} The smooth motion breaks as the inflection point reaches  the boundary of the barrier, $t=36$}\label{SL4}
\end{figure}

\begin{figure}[h]
\begin{minipage}{13.2pc}
\includegraphics[width=13.2pc]{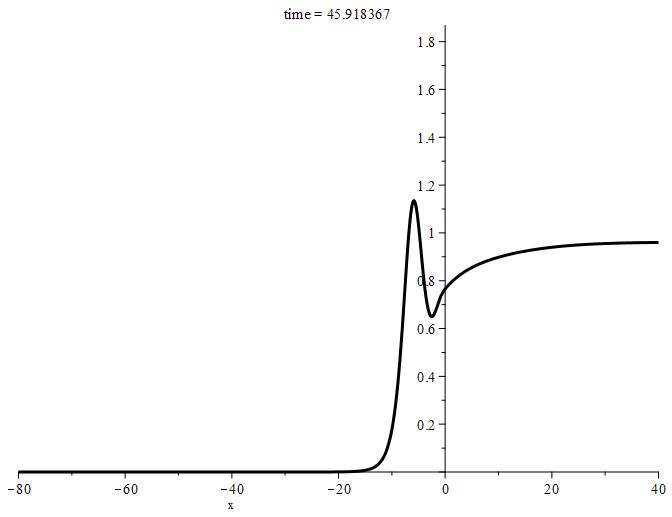}
\end{minipage}
\begin{minipage}{13.2pc}
\includegraphics[width=13.2pc]{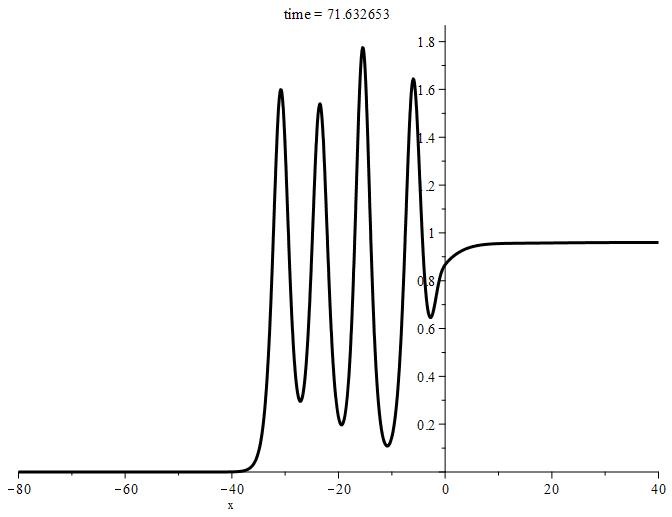}
\end{minipage}\caption{\textsl{\textbf{Left}:} Oscillation begins, $t=45$. \protect\newline\textsl{\textbf{Right}:} and continues $t=72$}\label{SL5}
\end{figure}

Next figures show that the quasi-harmonic oscillations develop, of a kind known for KdV, see figures \ref{SL4}--\ref{SL5}. You may also see the \textit{\textbf{inverse.avi }} Maple-generated  movie attached to this paper.

\section{Conclusion and numeric considerations}

The results may be of a practical use. For once, the form of the reflected wave may be used to estimate the thickness and/or the density of the viscous barrier. A refraction may also be predicted.

The figures in this paper were generated numerically using Maple PDETools package. The  mode of operation uses the default Euler method, which is a centered implicit scheme, and  can be used to find solutions to evolution PDEs. This implicit scheme is unconditionally stable for many problems (though this may need to be checked).

 Yet note that an accurate presentation of oscillations and/or sole peaks requires the choice of the Maple procedure's \emph{spacestep} and/or the \emph{timestep} parameters corresponding to a typical length and height of the solution detail.

 Qualitative estimations of the refraction coefficient, based on the relative decay of the KdV selected conservation laws will be published elsewhere.

\end{document}